\documentclass{elsart}
\usepackage{graphicx}

\usepackage{amssymb}

\begin{document}

\begin{frontmatter}

\title{Spherical Couette flow in a dipolar magnetic field}
%shorttitle{Magnetic spherical Couette flow}

\author{Rainer Hollerbach$^{\rm a}$,}
\author{Elisabeth Canet$^{\rm b}$,}
\author{Alexandre Fournier$^{\rm b}$}

\address[add1]{Department of Applied Mathematics, University of Leeds,
               Leeds, LS2 9JT, United Kingdom}
\address[add2]{Laboratoire de G\'eophysique Interne et Tectonophysique,
               Universit\'e Joseph Fourier, BP 53, 38041 Grenoble
               Cedex 9, France}

\begin{abstract}
We consider numerically the flow of an electrically conducting fluid in
a differentially rotating spherical shell, in a dipolar magnetic field.
For infinitesimal differential rotation the flow consists of a
super-rotating region, concentrated on the particular field line
$\mathcal C$ just touching the outer sphere, in agreement with previous
results.  Finite differential rotation suppresses this super-rotation,
and pushes it inward, toward the equator of the inner sphere.  For
sufficiently strong differential rotation the outer boundary layer
becomes unstable, yielding time-dependent solutions.  Adding an
overall rotation suppresses these instabilities again.  The results
are in qualitative agreement with the DTS liquid sodium experiment.
\end{abstract}

\begin{keyword}
Magnetohydrodynamics \sep Spherical Couette flow
\PACS 47.20.Qr \sep 47.65.-d
\end{keyword}
\end{frontmatter}

\section{Introduction}

Spherical Couette flow is the flow induced in a spherical shell by
differentially rotating the inner and/or outer spheres.  Despite its
simplicity, this configuration yields a broad range of different flow
patterns, which form an important part of classical fluid dynamics.
Now suppose that the fluid is electrically conducting, and a magnetic
field is imposed.  Magnetohydrodynamic effects then arise, which can
radically alter the flow structures from the previous nonmagnetic ones.
In this paper we present numerical solutions of spherical Couette flow
in a dipolar magnetic field, and compare them, at least qualitatively,
with the DTS ({\it Derviche Tourneur Sodium}) liquid sodium experiment
[1,2].

The mechanism whereby a magnetic field alters the flow is via the
magnetic tension force, coupling fluid along the magnetic field lines.
For a dipolar field, this singles out the particular field line
$\mathcal C$ just touching the outer sphere at the equator [3].  Fluid
inside $\mathcal C$ is coupled only to the inner sphere, and hence
co-rotates with it, whereas fluid outside $\mathcal C$ is coupled to
both spheres, and rotates at some intermediate rate.  The shear layer
on $\mathcal C$ separating the two regions scales as $Ha^{-1/2}$, where
the Hartmann number $Ha$ is a measure of the strength of the imposed
field.

However, this applies only if both boundaries are insulating.  If the
inner boundary is conducting, one obtains not a shear layer on $\mathcal
C$, but rather a super-rotating jet, that is, a region of fluid rotating
faster than either boundary [4,5].  The thickness of this jet still
scales as $Ha^{-1/2}$, and the amount of super-rotation saturates at
around 30\% of the imposed differential rotation.  Finally, if both
boundaries are conducting, the amount of super-rotation does not
saturate, but instead increases as $Ha^{1/2}$ [6,7].

So, one interesting aspect of the DTS experiment might be to study
this super-rotation, and how it depends on the various parameters in
the problem.  However, no clear evidence for super-rotation was found,
at least not on this particular field line $\mathcal C$.  This in turn
motivates us to reconsider this problem numerically, and attempt to
reconcile the experimental results with the previous theoretical
ones.  We will find that the crucial feature is the inertial term
$Re\,\bf U\cdot\nabla U$, which was not included in the previous
theoretical studies, but which is certainly important in the experiment.
Indeed, studying this regime where inertial effects can be as or even
more important than magnetic effects was one of the main motivations
in setting up the experiment.  We show here that including inertia
radically alters the flow patterns, and eliminates the special role
previously played by the field line $\mathcal C$.  For sufficiently
large Reynolds numbers we obtain instabilities similar to ones found in
the experiment.  Finally, we show how an overall rotation suppresses
these instabilities again, also in qualitative agreement with the
experimental results.

\section{Equations}
In a reference frame co-rotating with the outer sphere, the
suitably nondimensionalized equations are
$$\frac{\partial{\bf U}}{\partial t} + Re\,{\bf U\cdot\nabla U}
  + 2E^{-1}\,{\bf\hat e}_z\times{\bf U}$$
$$ = -\nabla p + \nabla^2{\bf U}
  + Ha^2\,Rm^{-1}\,(\nabla\times{\bf B})\times{\bf B},\eqno(1)$$

$$Re^{-1}\,\frac{\partial{\bf B}}{\partial t}=Rm^{-1}\,\nabla^2{\bf B}
  + \nabla\times({\bf U\times B}),\eqno(2)$$
together with $\nabla\cdot{\bf U}=0$ and $\nabla\cdot{\bf B}=0$.
Length has been scaled by the outer sphere's radius $r_o$, time by the
viscous diffusive timescale $r_o^2/\nu$, and the flow $\bf U$ by
$r_i\Delta\Omega$, where $r_i$ is the inner sphere's radius, and
$\Delta\Omega=\Omega_i-\Omega_o$ is the differential rotation between
the inner and outer spheres.  Note incidentally how the flow scale is
not given by lengthscale/timescale.  The scalings adopted here were
chosen because they include the $\Delta\Omega\to0$ limit of infinitesimal
differential rotation in a particularly convenient form, simply as $Re
\to0$ in (7).

The Hartmann number
$$Ha=\frac{B_0r_o}{\sqrt{\mu\rho\nu\eta}},\eqno(3)$$
where $\mu$ is the permeability, $\rho$ the density, $\nu$ the kinematic
viscosity, and $\eta$ the magnetic diffusivity, is a measure of the
strength $B_0$ of the imposed dipole field, at $r=1$, $\theta=\pi/2$
(where $r,\theta,\phi$ are standard spherical coordinates).  $B_0$ is
then also the scaling for the field $\bf B$.

The inverse Ekman number
$$E^{-1}=\frac{\Omega_o\,r_o^2}{\nu}\eqno(4)$$
measures the outer sphere's rotation rate, and the two Reynolds numbers
$$Re=\frac{\Delta\Omega\,r_ir_o}{\nu},\qquad
  Rm=\frac{\Delta\Omega\,r_ir_o}{\eta}\eqno(5)$$
both measure the differential rotation rate, compared with the viscous
and magnetic diffusive timescales, respectively.

The ratio $Rm/Re=\nu/\eta$ is a material property of the fluid, referred
to as the magnetic Prandtl number $Pm$.  For liquid sodium $Pm\sim10^{-5}$.
This extremely small value of $Pm$ means that $Re$ can be quite large while
$Rm$ is still small.  This allows us to further simplify the governing
equations in the following way: Expand the field as
$${\bf B}={\bf B}_0 + Rm\,{\bf b},\eqno(6)$$
where ${\bf B}_0=2\cos\theta\,r^{-3}\,{\bf\hat e}_r
+ \sin\theta\,r^{-3}\,{\bf\hat e}_\theta$ is the imposed dipole field
(now normalized to $|{\bf B}_0|=1$ at $r=1$, $\theta=\pi/2$), and
$Rm\,{\bf b}$ is the induced field.  Inserting (6) into (1-2) and
neglecting $O(Rm)$ terms then yields
$$\frac{\partial{\bf U}}{\partial t} + Re\,{\bf U\cdot\nabla U}
  + 2E^{-1}\,{\bf\hat e}_z\times{\bf U}$$
$$ = -\nabla p + \nabla^2{\bf U}
  + Ha^2\,(\nabla\times{\bf b})\times{\bf B}_0,\eqno(7)$$

$$\nabla^2{\bf b}=-\nabla\times({\bf U\times B}_0),\eqno(8)$$
which (by construction) no longer involves $Rm$ at all.  That is, $Rm$
enters only in the meaning we ascribe to $\bf b$, but not in the equations
we actually solve.  The advantage of this small $Rm$ approximation is that
the induction equation (8) is no longer time-stepped at all, but simply
inverted at each time-step of the momentum equation (7).  Filtering out
the magnetic diffusive timescale in this way then allows for larger
time-steps than would otherwise be possible.

The boundary conditions associated with (7) are
$${\bf U}=r\sin\theta\,{\bf\hat e}_\phi\quad{\rm at}\quad r=r_i,
  \qquad{\bf U}=0\quad{\rm at}\quad r=r_o.\eqno(9)$$
Those associated with (8) are a little more complicated.  We begin by
decomposing $\bf b$ as
$${\bf b}=      \nabla\times(g\,{\bf\hat e}_r)
  + \nabla\times\nabla\times(h\,{\bf\hat e}_r),\eqno(10)$$
and expanding $g$ and $h$ in Legendre polynomials
$$g=\sum_l g_l(r,t)\,P_l(\cos\theta),\qquad
  h=\sum_l h_l(r,t)\,P_l(\cos\theta).\eqno(11)$$
The boundary conditions at $r_i$ are then
$$\frac{d}{dr}\,g_l-\frac{l+1}{r}\,g_l=0,\qquad
  \frac{d}{dr}\,h_l-\frac{l+1}{r}\,h_l=0,\eqno(12)$$
corresponding to a conducting inner sphere [8].  The boundary conditions
at $r_o$ are
$$\epsilon\frac{d}{dr}\,\Bigl(rg_l\Bigr)+g_l=0,\qquad
  \frac{d}{dr}\,h_l+\frac{l}{r}\,h_l=0,\eqno(13)$$
corresponding to a thin outer layer of relative conductance $\epsilon$.
That is, $\epsilon=\delta\sigma_w/r_o\sigma_f$, where $\delta$
is the thickness of the outer wall, and $\sigma_w$ and $\sigma_f$ are
the conductivities of this wall and the fluid, respectively.  Taking
$\delta$ to be so small that the field can be assumed to vary only
linearly within the wall, one then obtains the boundary conditions (13).
(For finite $\delta$ the boundary conditions would be exactly the same,
except that $\epsilon$ would vary with $l$.)  In the DTS experiment
$\epsilon\approx0.003$; we will consider values between 0 and 1.  It
turns out though that with the inclusion of inertia the conductivity of
the outer boundary is less important than it was without it [6,7].

The equations (7,8), together with the boundary conditions (9,12,13),
were solved using the numerical code [9].  The radius ratio was fixed
at $r_i/r_o=1/2$.  The experimental value is actually 1/3 rather than
1/2, but the $r^{-3}$ dipole field strength then varies by more than
an order of magnitude across the shell, which would make $r_i/r_o=1/3$
numerically rather challenging.  By working in a somewhat thinner shell
one can reach higher values for $Ha$, while still obtaining qualitatively
much the same flow structures.  We also restrict attention to
axisymmetric solutions, which again allows us to reach higher parameter
values.  The experimental results are not purely axisymmetric, but they do
not show much non-axisymmetric structure, suggesting that axisymmetric
calculations are a reasonable first attempt at understanding them.

With these limitations of $r_i/r_o=1/2$ rather than 1/3, and 2D rather
than 3D calculations, we are able to reach parameter values as large
as $Ha=O(10^2)$, $Re=O(10^4)$ and $E^{-1}=10^6$.  For comparison, the
experimental values are $Ha=210$ (fixed, since the dipole embedded in
the inner sphere cannot be adjusted in strength), $Re=10^5-10^6$, and
$E^{-1}=10^6-10^8$ (if the outer sphere is rotating at all, otherwise
$E^{-1}=0$).  Smaller values of $Re$ and $E^{-1}$ are unfortunately not
attainable in the experiment, as the rotation rates of the inner and
outer spheres cannot be controlled accurately if they are too small.

Comparing these numbers, we see therefore that the numerically
achievable values are all somewhat smaller than the experimental ones,
but are sufficiently close that the numerical results might be expected
to give some qualitative insight at least into the experimental ones.
In particular, we are able to achieve the experimental values for the
additional two parameters $\Lambda=Ha^2E$, the Elsasser number measuring
magnetic versus Coriolis effects, and $N=Ha^2/Re$, the so-called
interaction parameter measuring magnetic versus inertial effects.  We
will see though that $N$ in particular is not necessarily the most
relevant comparison of $Ha$ and $Re$; other ratios such as $Re\sim Ha$
or $Re\sim Ha^{0.7-0.8}$ seem to play a more important role.

\section{Results}

Figure 1 shows the angular velocity for $E^{-1}=0$ (no overall rotation),
$Re=0$ (infinitesimal differential rotation), and $Ha^2=10^3$ to $10^6$
(for $Re=0$ the problem is linear, so one can achieve far larger values
of $Ha$ than for $Re>0$).  The top row has $\epsilon=0$, so an insulating
outer boundary; the bottom row has $\epsilon=1$, a strongly conducting
outer boundary.  In both cases we obtain precisely the results mentioned
in the introduction: with increasing $Ha$ the flow is increasingly
concentrated on the field line $\mathcal C$, and the degree of
super-rotation levels off at 36\% for $\epsilon=0$, but increases
monotonically for $\epsilon=1$.

\begin{figure}
\centering
\includegraphics[width=1.0\textwidth]{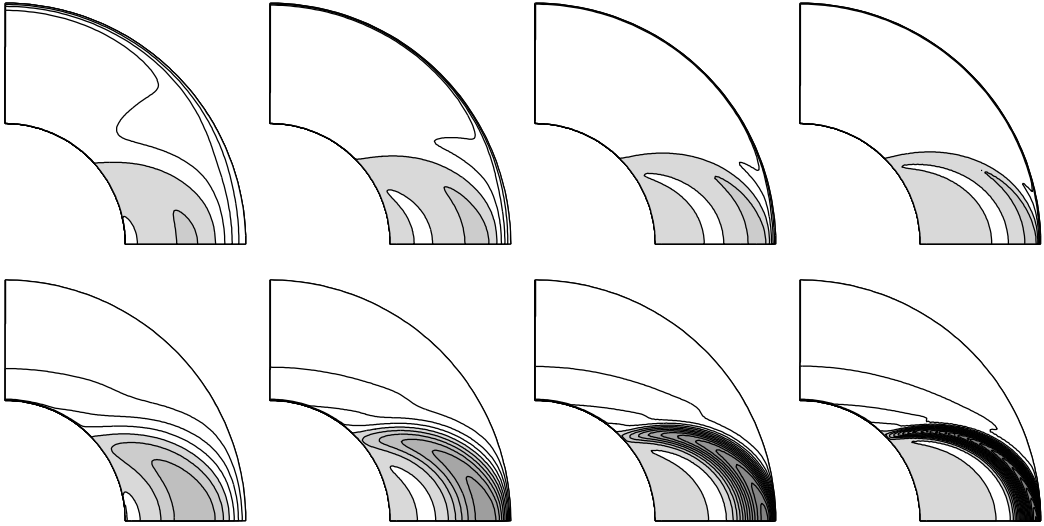}
\caption{Contour plots of the angular velocity for $\epsilon=0$ (top row)
and $\epsilon=1$ (bottom row).  From left to right $Ha^2=10^3$, $10^4$,
$10^5$ and $10^6$.  $Re=E^{-1}=0$.  Contour interval of 0.25, with
super-rotating regions gray-shaded.  The maximum values are 1.28, 1.32,
1.35 and 1.36 in the top row, and 1.71, 2.45, 3.89 and 6.54 in the
bottom row.}
\end{figure}

The left panel in Fig.\ 2 quantifies how the super-rotation varies with
$Ha$, for different values of $\epsilon$.  For $\epsilon=1$ it does indeed
appear to scale as $Ha^{1/2}$, as predicted by the asymptotic analysis [7].
Turning next to $\epsilon=0.1$, $10^{-2}$ and $10^{-3}$, we note that for
sufficiently large $Ha$ even $\epsilon=10^{-3}$ deviates from $\epsilon=0$,
and starts to rise.  This would seem to confirm the suggestion made by [8]
that the relevant ratio is not the boundary's conductance compared with the
conductance of the entire depth of fluid, but only with the conductance of
the Hartmann boundary layer.  The thickness of this layer scales as $Ha^{-1}$,
suggesting that the relevant parameter is not $\epsilon$ itself, but rather
$\epsilon\,Ha$.  Once this exceeds $O(1)$, the boundary is qualitatively more
like conducting than insulating.  For any non-zero $\epsilon$ the degree of
super-rotation should therefore eventually start to rise, just as seen here.

\begin{figure}
\centering
\includegraphics[width=1.0\textwidth]{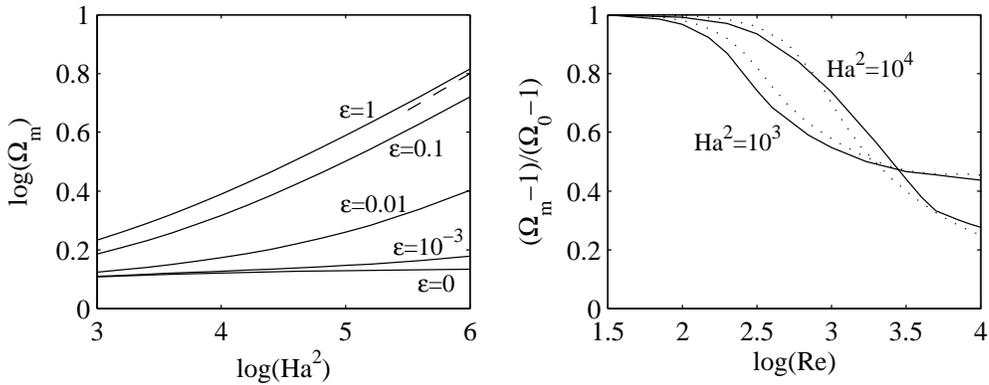}
\caption{The left panel shows how the maximum value of the angular
velocity varies with $Ha$, for the indicated values of $\epsilon$.
The dashed line segment alongside the $\epsilon=1$ curve denotes the
$Ha^{1/2}$ expected asymptotic scaling.  The right panel shows how the
quantity $(\Omega_m-1)/(\Omega_0-1)$ varies with $Re$, where $\Omega_m$
is the maximum value of the angular velocity at the given $Re$, and
$\Omega_0$ is the maximum value of the angular velocity at $Re=0$.
Solid lines are $\epsilon=1$, dotted lines are $\epsilon=0$, and
Hartmann numbers as indicated.}
\end{figure}

All these results so far have been for $Re=0$, the infinitesimal
differential rotation limit considered previously [3-7].  Figure 3 shows
how the flow is altered if we now increase $Re$, to the point where
inertial and magnetic effects are comparable (that is, $N=1$).  We note
first that in addition to the angular velocity, there is now a meridional
circulation as well, which has a significant effect back on the angular
velocity.  In particular, the previous super-rotation on $\mathcal C$ is
strongly suppressed, and also pushed inward, until there is nothing left
of the original structure on $\mathcal C$.  Another interesting feature
to note is how the meridional circulation compresses the remaining outer
boundary layer until it is much thinner than the original, linear
boundary layer.

\begin{figure}
\centering
\includegraphics[width=1.0\textwidth]{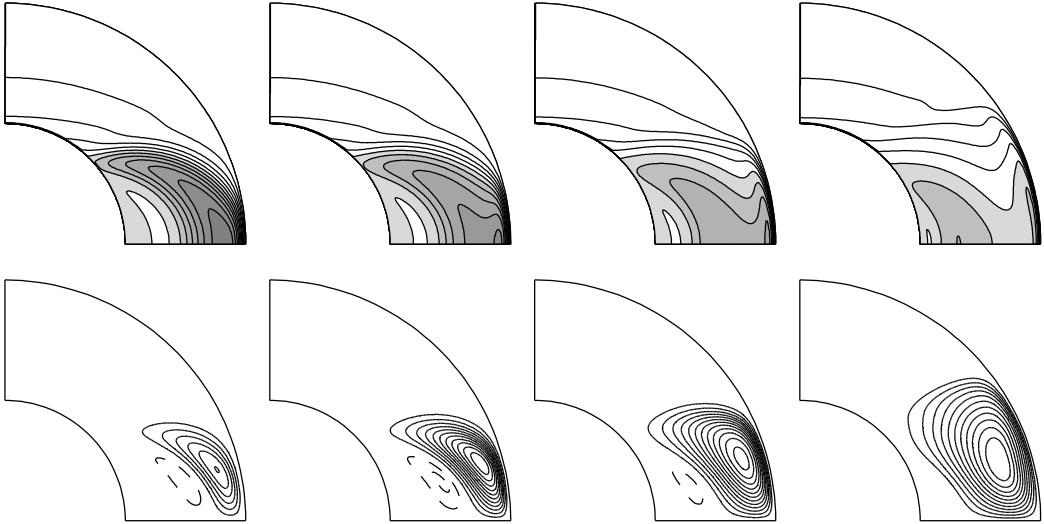}
\caption{$Ha^2=10^4$, $E^{-1}=0$, $\epsilon=1$, and from left to right
$Re=300$, 1000, 3000 and 10000.  The top row shows contours of the angular
velocity, with a contour interval of 0.2, and super-rotating regions
gray-shaded.  The bottom row shows streamlines of the meridional
circulation, with solid lines denoting counter-clockwise circulation (with
a contour interval of $10^{-3}$), and dashed lines denoting a much weaker
clockwise circulation (with a contour interval of $5\cdot10^{-4}$).  The
angular velocity is symmetric about the equator, the meridional
circulation anti-symmetric.  Perturbations of the opposite symmetry were
introduced, but decayed away in every case.}
\end{figure}

This behavior is very different from that obtained for an axial rather
than a dipole field, where the axisymmetric basic state is almost
unaffected by increasingly large $Re$, right up to the onset of
non-axisymmetric instabilities [8].  The difference is that for a uniform
axial field the flow is correspondingly also largely independent of $z$.
If $\bf U$ only depends on $s$ though (where $z,s,\phi$ are cylindrical
coordinates), then so does $\bf U\cdot\nabla U$, which can therefore be
balanced by $-\nabla p$.  In contrast, for the non-uniform dipole field
considered here, $\bf U$ clearly depends on both coordinates $r$ and
$\theta$, so $\bf U\cdot\nabla U$ cannot so easily be balanced by
$-\nabla p$, but instead fundamentally alters the flow, as we see in
Fig.\ 3.

Returning to Fig.\ 2, the right panel quantifies the suppression of the
super-rotation, showing how $(\Omega_m-1)/(\Omega_0-1)$ varies with $Re$,
where $\Omega_m$ is again the maximum angular velocity, at the given $Re$,
and $\Omega_0$ is the maximum angular velocity at $Re=0$.  That is, this
quantity measures the relative amount by which the original super-rotation
has been suppressed.  Solid lines denote $\epsilon=1$, dotted lines
$\epsilon=0$.  We see therefore that even though conducting versus
insulating outer boundaries yield very different degrees of
super-rotation, the relative extent to which it is suppressed by $Re$
is surprisingly similar.

Note also how larger Hartmann numbers require larger $Re$ before the
super-rotation starts to get suppressed.  For example, if we focus on how
large $Re$ must be before the super-rotation is suppressed to 80\% of its
original value, we find that it must be some 3 times larger for $Ha^2=10^4$
than for $Ha^2=10^3$, perhaps suggesting an $Re\sim Ha$ scaling.  If true,
this would indicate that the interaction parameter $N$ is not in fact the
most appropriate measure of magnetic versus inertial effects for this
problem.

The last point to note in this panel of Fig.\ 2 is that while larger
Hartmann numbers may also require larger Reynolds numbers before the
super-rotation starts to get suppressed, for sufficiently large $Re$ it is
actually suppressed more for larger $Ha$.  In particular, this suggests
that for $Ha=210$ and $Re\gtrsim10^5$ it might be suppressed to perhaps no
more than 10\% of its original $Re=0$ value, which would be consistent with
the experimental findings of no clearly detectable super-rotation at all.
(However, we must remember also that what little super-rotation is left is
no longer situated on $\mathcal C$, but is instead concentrated at the
inner sphere, where the experiment currently cannot make measurements.
Once further ultrasound transducers are installed to measure the flow
closer to the inner sphere, according to our results here they should
measure at least some slight super-rotation.)

Figure 4 shows the solution at $Ha^2=10^3$ and $Re=10^4$.  In
addition to the remaining slight super-rotation at the equator of the inner
sphere, there is now a new feature, namely a time-dependence near the outer
sphere.  The outer boundary layer periodically breaks down in mid-latitudes
into a series of small-scale ripples, with period 0.0060, or $120\,
\Omega_i^{-1}$ on the rotational timescale.  None of this time-dependence 
penetrates very far into the interior though.  One possible
explanation for this is simply the $r^{-3}$ dipole field strength, which
increases so strongly going inward that the interior is still magnetically
dominated even at these values of $Re$.

\begin{figure}
\centering
\includegraphics[width=1.0\textwidth]{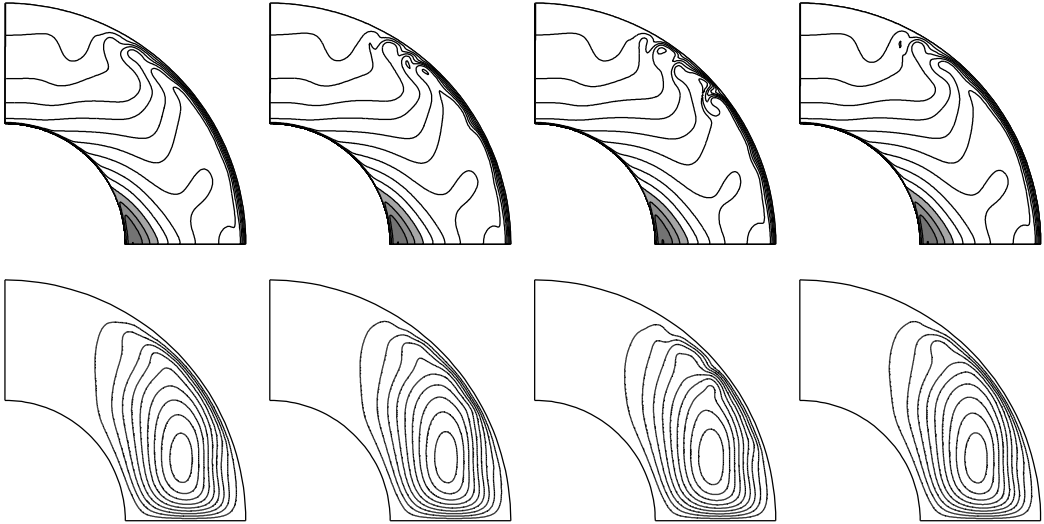}
\caption{The solution at $Ha^2=10^3$, $Re=10^4$, $E^{-1}=0$, and $\epsilon
=1$.  From left to right four snapshots of the time-dependent solution,
uniformly spaced throughout the period 0.0060, or $120\,\Omega_i^{-1}$.  The
top row shows contours of the angular velocity, with a contour interval of
0.1.  The bottom row shows streamlines of the meridional circulation, with
a contour interval of $10^{-3}$.  Perturbations of the opposite equatorial
symmetry were introduced, but decayed away.}
\end{figure}

Increasing $Re$ further, Fig.\ 5 shows the solution at $Re=15000$.  The
boundary layer eruptions are now considerably more pronounced, and cover a
much broader range in latitude, including near the equator of the outer
sphere.  They are also no longer equatorially symmetric, but instead
alternate between the two hemispheres.  The period is 0.0014, or $42\,
\Omega_i^{-1}$.  This basic periodicity is quite regular, but the details
of the individual pulses are not; the solution is evidently quasi-periodic.

\begin{figure}
\centering
\includegraphics[width=1.0\textwidth]{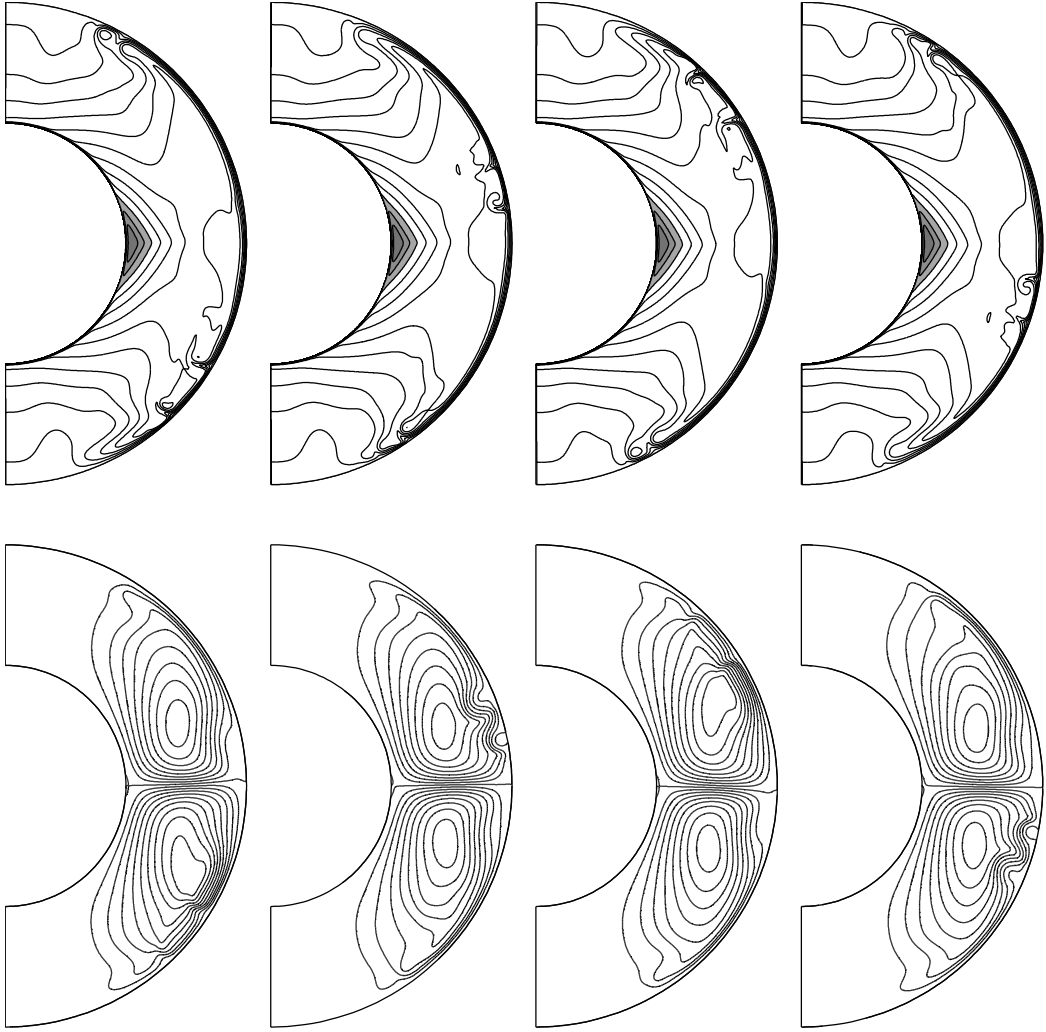}
\caption{As in Fig.\ 4, but at $Re=15000$.  The period is 0.0014, or
$42\,\Omega_i^{-1}$.  The numerical resolution was 135 Chebyshev polynomials
in $r$ times 720 Legendre polynomials in $\theta$, and was checked to ensure
that even these very fine structures are adequately resolved.}
\end{figure}

We note that the experiment also exhibits rapid
fluctuations near the outer boundary, but a much more quiescent interior.
To assess whether this might be related to our results here, we need to
know how the critical Reynolds number for the onset of this time-dependence
scales with $Ha$.  Increasing $Re$ in steps of 200, we obtained $Re_c=9800$,
12600 and 16400, for $Ha^2=1000$, 2000 and 4000, respectively.  The ratios
12600/9800=1.29 and 16400/12600=1.30 then suggest the scaling $Re_c\sim
Ha^{0.74}$, although of course with only three data points, spanning a
range of just 2 in $Ha$, one should not assign too much significance to
this precise exponent 0.74.  Nevertheless, it again demonstrates that the
interaction parameter $N$ is not necessarily the most relevant ratio of
Hartmann and Reynolds numbers.  Furthermore, it suggests that the experiment
should be far above the critical Reynolds number for the onset of this
time-dependence, so the fluctuations observed in the experiment may indeed
include these instabilities discovered here (in addition to possible
non-axisymmetric instabilities not considered here).

It is interesting also to compare our $Re_c\approx760Ha^{0.74}$
instabilities with the Hartmann layer instabilities explored by
[10,11], who found that $Re_c\approx380Ha$.  Inserting our values of $Re_c$
and $Ha$, for $Ha^2=1000$, 2000 and 4000 we obtain $Re_c/Ha=310$, 280 and
260; sufficiently close to 380 that the instabilities are likely to be
closely related.  The slightly different scalings with $Ha$ could plausibly
be explained by the fact that the spherical shell geometry considered here
is considerably more complicated than the planar geometry considered by
[10,11], and correspondingly our basic state depends on $Ha$ and $Re$ in a
more complicated way than it does in planar geometry.  See also [12], who
consider Hartmann layer instabilities in a rather different parameter
regime, appropriate to the Earth's rapidly rotating core.

Finally, the last point to note is that $\epsilon=0$ yields solutions
similar to Figs.\ 4 and 5, merely at somewhat larger values of $Re$.  This
is perhaps not surprising: the main effect of $\epsilon$ appears to be to
control the degree of super-rotation, just as it did in the linear regime,
but as Figs.\ 4 and 5 show, this time-dependence is completely unrelated
to the remaining, rather weak super-rotation.

All results so far have been for $E^{-1}=0$, so a stationary outer sphere.
In the experiment it was found that rotating the outer sphere tended to
suppress these fluctuations near the outer boundary.  We would therefore
like to test whether adding an overall rotation will similarly suppress
our instabilities in Figs.\ 4 and 5.  But first, Fig.\ 6 shows the effect
of adding a non-zero $E^{-1}$ to the previous solution in Fig.\ 3.  Not
surprisingly, an increasingly rapid overall rotation eventually suppresses
all the previous structure, and the flow becomes almost completely aligned
with the $z$-axis.  Similar solutions were also obtained by [13], but
coming from a rather different direction in parameter space, namely
starting with a rapid rotation, and seeing how an increasingly strong
magnetic field suppresses the so-called Stewartson layer on the tangent
cylinder.

\begin{figure}
\centering
\includegraphics[width=1.0\textwidth]{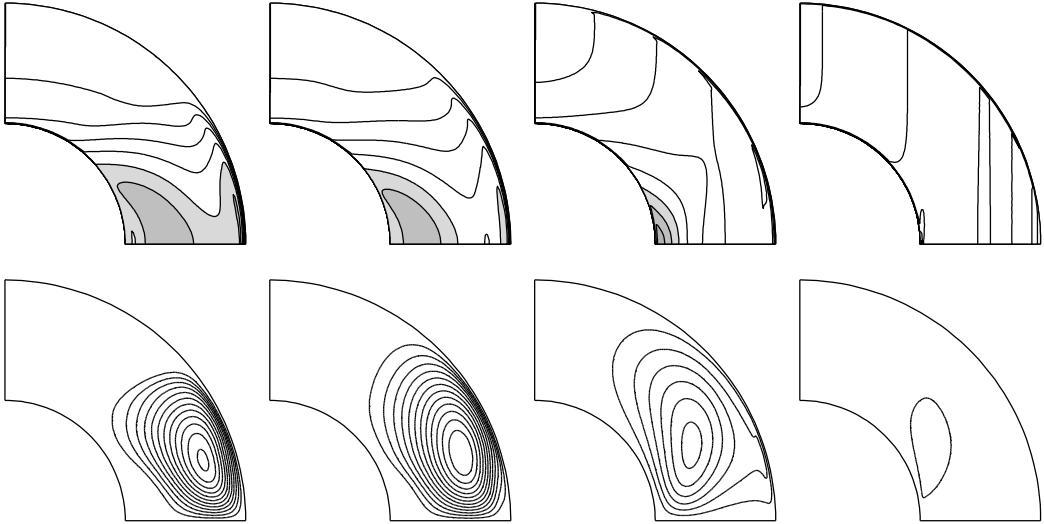}
\caption{$Ha^2=10^4$, $Re=10^4$, $\epsilon=1$, and from left to right
$E^{-1}=10^3$, $10^4$, $10^5$ and $10^6$.  The top row shows contours of
the angular velocity, with a contour interval of 0.2; the bottom row shows
streamlines of the meridional circulation, with a contour interval of
$10^{-3}$.}
\end{figure}

Given how effectively the Coriolis force suppresses the previous structure,
it seems likely that it will also suppress the instabilities.  Figure 7
shows that this is indeed the case; one can increase $Re$ up to 25000 at
least, and still finds nothing like the instabilities in Figs.\ 4 and 5.
One other interesting point to note regarding Figs.\ 6 and 7 is that the
$Ha^2=10^4,\ E^{-1}=10^5$ and $Ha^2=10^3,\ E^{-1}=10^4$ solutions look
rather similar, and similarly $E^{-1}=10^6$ and $E^{-1}=10^5$.  This
indicates that, unlike the interaction parameter $N$, the Elsasser number
$\Lambda=Ha^2E$ is indeed the relevant parameter here.

\begin{figure}
\centering
\includegraphics[width=1.0\textwidth]{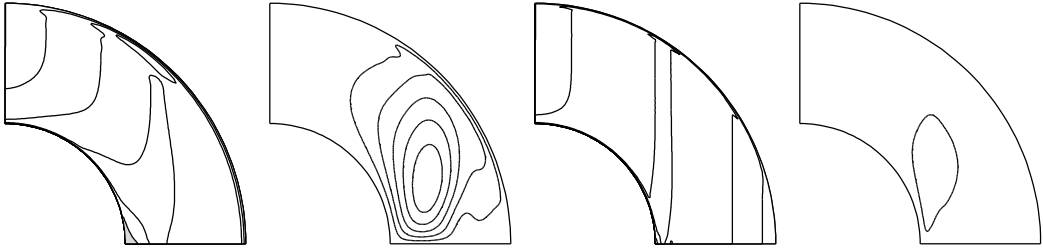}
\caption{$Ha^2=10^3$, $Re=25000$, $\epsilon=1$.  The first two panels
show the angular velocity and meridional circulation at $E^{-1}=10^4$,
the second two panels at $E^{-1}=10^5$.  Contour intervals again 0.2
and $10^{-3}$.}
\end{figure}

\section{Conclusion}

We have found that the inclusion of inertia in this magnetic spherical
Couette flow problem radically alters the results, suppressing the
previous super-rotation, and completely eliminating the significance of
the field line $\mathcal C$.  For sufficiently large Reynolds numbers we
also discovered instabilities in the outer boundary layer, which may be
related to some of the fluctuations seen in the experiment, particularly
as a rapid overall rotation suppresses them again in both the experiment
and here.

Finally, there are (at least) two further issues that should be explored
numerically.  First, what about non-axisymmetric instabilities, for
example G\"ortler vortices associated with the meridional circulation?
It would certainly be of interest to compute critical Reynolds numbers
for their onset, and see whether they are greater or smaller than the
$Re_c\approx760Ha^{0.74}$ onset of the axisymmetric instabilities
considered here.  The experiment did not show any large-scale
non-axisymmetric structures, but was not purely axisymmetric either.
This suggests that the most unstable non-axisymmetric modes may have very
high azimuthal wavenumber $m$ (which would be consistent with small-scale
G\"ortler vortices).  If the 3D solutions exhibit structure in $\phi$
comparable to the structure in $\theta$ seen in Figs.\ 4 and 5, that
would certainly correspond to very high $m$ indeed, making fully 3D
solutions very difficult.

Second, we recall that all of our calculations were in the small $Rm$
limit (7,8).  For the solutions here, this is appropriate, since $Re=
25000$, the largest value considered, still corresponds to $Rm<1$.  In
the experiment though $Re$ is so large that even $Rm>1$.  Furthermore,
finite $Rm$ opens up the possibility of fundamentally new dynamics,
such as the magnetorotational instability.  It would be of considerable
interest therefore to consider finite $Rm$, and see whether anything
emerges that is completely different from the results presented here.
Some of these calculations are currently under way.

\section*{Acknowledgment}
We thank Thierry Alboussi\`ere, Daniel Brito, Philippe Cardin,
Dominique Jault, Henri-Claude Nataf and Denys Schmitt for stimulating
discussions on the DTS experiment.

\end{document}